\documentclass[prd,twocolumn,superscriptaddress,altaffilletter,nofootinbib]{revtex4}

\usepackage{cancel}
\usepackage{graphicx}
\usepackage{xcolor}
\usepackage{amsmath}
\usepackage{amssymb}
\usepackage{graphicx,epsfig}
\newcommand{\be}{\begin{equation}}
\newcommand{\ee}{\end{equation}}
\newcommand{\bea}{\begin{eqnarray}}
\newcommand{\eea}{\end{eqnarray}}
\newcommand{\der}{\partial}

\begin{document}

%%%%%%%%%%%%%%%%%%%%%%%
%%%%%%%%%%%%%%%%%%%%%%%

\title{Local scale symmetry in non-riemannian geometry based gravitational theories and the role of the noether current}

\author{R. Gonzalez Quaglia}\email{rodrigo@icf.unam.mx}\affiliation{Instituto de Ciencias F\'isicas, Universidad Nacional Aut\'onoma de M\'exico, Av. Universidad S/N. Cuernavaca, Morelos, 62251, M\'exico.}

\author{Israel Quiros}\email{iquiros@fisica.ugto.mx}\affiliation{Departamento de Ingenier\'ia Civil, Divisi\'on de Ingenier\'ias, Universidad de Guanajuato, Campus Guanajuato, Guanajuato, 36000, M\'exico.}

%%%%%%%%%%%%%%%

\date{\today}

%%%%%%%%%%%%%%%%%%%%%%%%%%%%%%%%%%%%%%%%%%%%%%%%%%%%%%%%%%%%%%

\begin{abstract} In this paper, we delve into the significance of local scale symmetry and the role of the associated noether current, within gravitational theories which are based in non-riemannian background space. Our focus is in Weyl and in Riemann-Cartan geometry based gravitational theories. We show that local scale symmetry is associated with vanishing noether current whenever there are not new propagating gravitational degrees of freedom beyond the two polarizations of the massless graviton. In contrast, in local scale invariant theories where there are more than two propagating gravitational degrees of freedom, local scale symmetry is associated with nonvanishing noether current. The known result that Weyl symmetry has vanishing noether current, is generalized to non-riemannian gravitational theories. An exception are the local scale invariant gravitational theories with vectorial nonmetricity, where the associated noether current is nonvanishing.\end{abstract}

%%%%%%%%%%%%%%%%%%%%%%%%%%%%%%%%%%%%%%%%%%%%%%%%%%%%%%%%%%%%%

%\pacs{04.50.Kd, 04.50.Cd, 11.10.Ef, 98.80.-k, 98.80.Jk}

%%%%%%%%%%%%%%%%%%%%%

\maketitle

%%%%%%%%%%%%%%%%%%%%%

%%%%%%%%%%%%%%%%%%%%%%%%%%%%%%%%%%%%%%%%%%

\section{Introduction}\label{sect-intro}

%%%%%%%%%%%%%%%%%%%%%%%%%%%%%%%%%%%%%%%%%%

In recent times, scale invariance in the gravitational sector has gained significance as it has been motivated by pure gravitational and cosmological arguments \cite{kallosh-2013, kallosh-2013-1, bars-2014, bars-2014-1, bose-2016, ferreira-2017, ferreira-2019, ghile-2022, ghile-2023, harko-2023, quiros-2023-1, quiros-arxiv-23, jackiw-2015, garay, oda-2022}. In the context of cosmology, motivation arises from the success of the Starobinsky model \cite{Starobinsky-1980} as one of the most promising inflationary models \cite{guth-1981, linde-1982, albrecht-1982}. The Starobinsky model  incorporates a higher order curvature term, namely an $R^{2}$ term, together with the Einstein-Hilbert action. Indeed, this additional term drives the accelerated expansion during the early Universe.
Since inflation is believed to occur at high energies, the quadratic term in the Ricci scalar, being power counting scale invariant, becomes crucial as it suggests the existence of a gravitational scale-invariant theory. Moreover, it has been shown that this symmetry can be dynamically broken leading to the restoration of General Relativity and the generation of the Planck mass at the end of inflation \cite{ferreira-2019, Gialamas-2021} smoothly transitioning to the Big Bang universe. Moreover, the consideration of such a symmetry has also resulted in promising results in particle theory where it has been employed as a possible explanation on why the Higgs boson's mass remains small in the presence of beyond the Standard Model physics \cite{ferreira-2021}. 

An interesting aspect of the rebirth of interest in local scale (LS) symmetry in gravitational theories, is that in several cases Riemann background spaces are replaced by Weyl geometry spaces, as the geometric substratum of the gravitational theories \cite{bars-2014, ghile-2022, ghile-2023, harko-2023, quiros-2023-1, quiros-arxiv-23}. This is interesting because Weyl geometry is a natural arena for LS symmetry \cite{weyl-1918, weyl-book, many-weyl-book, london-1927, dirac-1973, utiyama-1973, adler-book, maeder-1978}. In other cases Riemann geometry space $V_4$ is replaced by Riemann-Cartan space $U_4$ \cite{hehl, hehl-rev, string, hammond-rev, pereira, ferraro, pop}. In this case the role of local scale symmetry is less known. 

The main motivation for this paper stems from the results of Ref. \cite{jackiw-2015}, where the conformally coupled scalar (CCS) theory given by the gravitational Lagrangian:

\bea L_\text{ccs}=\frac{\sqrt{-g}}{2}\left[\frac{1}{6}\phi^2 R+(\der\phi)^2-\frac{\lambda}{4}\phi^4\right],\label{grav-lag}\eea over Riemann space is examined. Here $R$ is the curvature scalar of Riemann space, $\phi$ is a compensating scalar field and we have used the shorthand notation $(\der\phi)^2\equiv g^{\mu\nu}\der_\mu\phi\der_\nu\phi$. In Ref. \cite{jackiw-2015} it was found that the noether current associated with LS invariance in the CCS theory vanishes, which means that the corresponding symmetry does not have any dynamical role. This result was confirmed in \cite{garay} within the framework of gravitational theories which are invariant under transverse diffeomorphisms and Weyl transformations (WTDiff.) In Ref. \cite{oda-2022} it was further generalized to any Weyl invariant gravitational theory in four dimensional riemannian geometry. In the latter reference it has been shown that the Weyl transformation is non-dynamical in the sense that it does not contain the derivative term of the transformation parameter as opposed to the conventional gauge transformation. 

A distinctive feature of the result discussed above is that it was obtained under the assumption that the background geometric structure is riemannian. In the present paper we shall further pursue the above studies of LS symmetry and its associated noether current, to include modifications of Riemann geometry such as Weyl geometry and Riemann-Cartan spaces. Our aim is to find a connection between local scale symmetry and the noether current in gravitational theories over the mentioned non-riemannian geometric backgrounds, with the hope to generalize the results obtained in Ref. \cite{jackiw-2015, garay, oda-2022} to these cases. In the present work, in order to derive the noether current which is associated with the Weyl scale symmetry, we shall rely upon the formalism developed in Ref. \cite{ferreira-2017}.

%--------------------organization-----------------------------------

The paper has been organized in the following way. In Section \ref{sect-scale-inv}, we revisit the class of action $S_\text{ccs}=\int d^4x L_\text{ccs}$, over Riemann space, where a coupling between the compensating scalar and the curvature scalar of Riemann space is left undefined: i. e., in \eqref{grav-lag} we make the replacement $\phi^2 R\rightarrow\alpha\phi^2 R$, where $\alpha$ is a free coupling \cite{ferreira-2019}. This allows to determine the conditions under which the action exhibits LS invariance. Our analysis, which follows the methodology outlined in \cite{ferreira-2017, ferreira-2019}, confirms that the model possesses the aforementioned symmetry only when $\alpha=1$. The noether current for this particular value of the coupling $\alpha$ vanishes, which aligns with the result of Ref. \cite{jackiw-2015}. 

In Section \ref{sect-weyl-geom}, we explore the local scale invariance within the framework of Weyl geometry. We consider a generalization of \eqref{grav-lag} which couples to Weyl geometry spaces. In terms of LC quantities the resulting Lagrangian can be decomposed as: $\alpha L_\text{ccs}+L_\text{stuck}$, where $L_\text{ccs}$ is the CCS Lagrangian \eqref{grav-lag} and $L_\text{stuck}$ is the Stueckelberg Lagrangian. The resulting noether current depends explicitly on the free parameters of the theory. We discuss the possible cases which arise. 

Sections \ref{sect-ecg}, \ref{sect-ecartan} and \ref{sect-ginv-ech} are dedicated to extend the former study to the Riemann-Cartan geometry based gravitational theory. This case is less known so that, such an important issue as the possible (not unique) transformation properties of the torsion, is to be discussed. Another issue, which is discussed in Section \ref{sect-ecartan}, is related with the boundary terms in the Einstein-Cartan-Holst theory. Section \ref{sect-ginv-ech} contains the detailed analysis of the LS symmetry within a Scalar-Einstein-Cartan-Holst (SECH) theory, where we derive the noether current and discuss its role whenever the LS symmetry is present. A generalization of the CCS theory adapted to Riemann-Cartan space, is briefly exposed in Section \ref{sect-u4-y4}. 

Section \ref{sect-discu} provides a comprehensive discussion of the paper's findings, along with outlining potential avenues for future research. Finally, in Section \ref{sect-conclu}, in order to summarize the key points and contributions of the paper, brief concluding remarks are given. For completeness an appendix Section \ref{app-a} have been included, where the details of the derivation of Eq. \eqref{ecart-eom} are provided.

%%%%%%%%%%%%%%%%%%%%%%%%%%%%%%%%%%%%%%%%%%

%%%%%%%%%%%%%%%%%%%%%%%%%%%%%%%%%%%%%%%%%%%%%%%%%%%%%%%%%%%%%%%%%%%%%%%%

\section{Conformally coupled scalar and Local Scale Invariance}
\label{sect-scale-inv}

%%%%%%%%%%%%%%%%%%%%%%%%%%%%%%%%%%%%%%%%%%%%%%%%%%%%%%%%%%%%%%%%%%%%%%%%

Let us consider the following action:\footnote{Here we use the metric signature: $(-1,+1,+1,+1)$, so that the signs of the curvature scalar and of the kinetic energy density of the scalar field are reversed with respect to other works in the bibliography where the signature $(+1,-1,-1,-1)$, is chosen instead (see for instance Ref. \cite{ferreira-2017, ferreira-2019, ghile-2022, ghile-2023, harko-2023}.)}

\bea S=\int d^4x\sqrt{-g}\left[\frac{\alpha}{12}\phi^2R+\frac{1}{2}(\der\phi)^2-\frac{\lambda}{8}\phi^4\right],\label{action}\eea where $R$ is the Ricci scalar, $\alpha$ is a coupling constant and $\lambda$ is another free parameter. Due to our signature convention the kinetic energy density term of the scalar field $\phi$ in the above action has the wrong sign, making it a ghost field. However, due to LS symmetry $\phi$ is not a (propagating) physical degree of freedom (DOF) so that the sign of its kinetic energy density term is harmless. In regard to the signature, let us make the following remark. In the action (4) of Ref. \cite{ferreira-2017} the correct sign of the kinetic term is considered, i. e., in \eqref{action} the following replacement is made: $(\der\phi)^2\rightarrow-(\der\phi)^2$. This means that the action (4) of \cite{ferreira-2017} can be, at most, global scale but not local scale invariant, since, as shown below, the wrong sign is required for the unit $\sqrt{-g}[\phi^2R/6+(\der\phi)^2]$ to be invariant under the LS transformations:\footnote{Local scale transformations \eqref{weyl-t} are not diffeomorphisms since these leave the coordinates untouched. The LS transformations act on the fields exclusively.}

\bea g_{\mu\nu}\rightarrow\Omega^2(x)g_{\mu\nu},\;\phi\rightarrow\Omega^{-1}(x)\phi,\label{weyl-t}\eea where the positive smooth function $\Omega(x)$, is the conformal factor.

For $\alpha=1$ the action \eqref{action}, together with the derived equations of motion (EOM), are invariant under \eqref{weyl-t}. This class of action has been investigated in several contexts \cite{bars-2014, jackiw-2015, oda-2022, ferreira-2017, ferreira-2019}. Its vanishing variation with respect to the metric $\delta_{\bf g}S=0$, leads to the Einstein's EOM:

\begin{align} &\frac{\alpha}{6}\phi^2G_{\mu\nu}+\frac{3-\alpha}{3}\der_\mu\phi\der_\nu\phi-\frac{3-2\alpha}{6}g_{\mu\nu}(\der\phi)^2\nonumber\\
&-\frac{\alpha}{3}\phi\left(\nabla_\mu\nabla_\nu-g_{\mu\nu}\nabla^2\right)\phi+\frac{\lambda}{8}\phi^4g_{\mu\nu}=0,\label{einst-eom}\end{align} where, as usual $G_{\mu\nu}:=R_{\mu\nu}-g_{\mu\nu}R/2$ is the Einstein's tensor and we used the notation $\nabla^2\equiv g^{\mu\nu}\nabla_\mu\nabla_\nu$. Meanwhile, requiring vanishing variation of \eqref{action} with respect to the scalar field $\delta_\phi S=0$ one gets the scalar field's EOM:

\bea \nabla^2\phi-\frac{\alpha}{6}\phi R+\frac{\lambda}{2}\phi^3=0.\label{phi-eom}\eea The trace of the Einstein's EOM \eqref{einst-eom} is given by:

\bea \frac{\alpha}{6}\phi^2R+(1-\alpha)(\der\phi)^2-\alpha\phi\nabla^2\phi-\frac{\lambda}{2}\phi^4=0.\label{trace-eom}\eea By comparing \eqref{phi-eom} with \eqref{trace-eom} we get,

\begin{align} (1-\alpha)\nabla_\mu\left(\phi\nabla^\mu\phi\right)=0,\label{conserved-current}\end{align} Whenever $\alpha\neq 1$ the current conservation equation: $\nabla_\mu j_N^\mu=0$, takes place, where we define the current: 

\bea j_N^\mu:=(1-\alpha)\phi\nabla^\mu\phi=\frac{1-\alpha}{2}\nabla^\mu\phi^2.\label{j-def}\eea This coincides with the noether current which is independently obtained through noether variation $j_N^\mu=\delta S_g/\delta\der_\mu\epsilon$ \cite{ferreira-2017}, where we have redefined the conformal factor $\Omega\equiv e^\epsilon$ and $\delta g^{\mu\nu}=-2\epsilon(x)g^{\mu\nu}$, $\delta\phi=-\epsilon(x)\phi$. For $\alpha=1$ the current \eqref{j-def} vanishes so that the current conservation equation \eqref{conserved-current} becomes an identity $0=0$. In this case ($\alpha=1$) the action \eqref{action} is invariant under the LS transformations \eqref{weyl-t} as we shall immediately show.\footnote{Under the Weyl transformations \eqref{weyl-t} the quartic term $\lambda\phi^4/4$ is obviously a LS invariant term so that sometimes, for simplicity, we shall omit it.} It suffices to show how does the unit:

\bea \sqrt{-g}\left[\frac{\alpha}{12}\phi^2R+\frac{1}{2}(\der\phi)^2\right],\label{unit}\eea transform under LS transformations \eqref{weyl-t} for arbitrary $\alpha$. We have that, under \eqref{weyl-t}: 

\begin{align} &R\rightarrow\Omega^{-2}\left(R-6\der_\mu\ln\Omega\der^\mu\ln\Omega-6\nabla^2\ln\Omega\right),\nonumber\\
&(\der\phi)^2\rightarrow\Omega^{-4}\left[(\der\phi)^2-\der_\mu\phi^2\der^\mu\ln\Omega\right.\nonumber\\
&\left.\;\;\;\;\;\;\;\;\;\;\;\;\;+\phi^2\der_\mu\ln\Omega\der^\mu\ln\Omega\right].\label{usef-2}\end{align} Taking into account \eqref{usef-2}, we can show that under \eqref{weyl-t}, the unit $(\ref{unit})$ transforms as:

\begin{align} &\sqrt{-g}\left[\frac{\alpha}{12}\phi^2R+\frac{1}{2}(\der\phi)^2\right]\rightarrow\sqrt{-g}\left[\frac{\alpha}{12}\phi^2R+\frac{1}{2}(\der\phi)^2\right.\nonumber\\
&\left.+\frac{1-\alpha}{2}\phi^2\der_\mu\ln\Omega\der^\mu\ln\Omega+\frac{1-\alpha}{2}\phi^2\nabla^2\ln\Omega\right],\nonumber\end{align} where the total derivative $-\nabla_\mu\left(\phi^2\der^\mu\ln\Omega\right)/2$, within square brackets in the right-hand side (RHS) of this equation, has been omitted since, within the action integral \eqref{action}, it amounts to a boundary term. Only for $\alpha=1$, the unit \eqref{unit} is invariant under the Weyl transformations \eqref{weyl-t}. Hence, $\alpha=1$ is required for local scale invariance to be a symmetry of the action \eqref{action}. However, for this specific value of the coupling constant $\alpha$, the noether current vanishes. Consequently, the necessary and sufficient requirement for LS invariance to be a symmetry of the gravitational theory \eqref{action}, is that the noether current associated with this symmetry vanished. 

Another way to understand why the requirement of vanishing noether current is necessary for LS invariance to be a a symmetry of \eqref{action} is through the analysis of the resulting EOM. For $\alpha=1$ the action \eqref{action} is LS invariant, as well as the derived equations of motion \eqref{einst-eom} and \eqref{phi-eom}, both evaluated at $\alpha=1$. In this case the trace of the Einstein's equation \eqref{trace-eom} coincides with the scalar field's EOM \eqref{phi-eom}, so that it is not an independent EOM. In consequence, the scalar field does not obey an independent equation of motion, i. e., $\phi$ is not a physical DOF. As previously remarked, the fact that the scalar field is a ghost field is not important since the requirement of LS invariance of \eqref{action} automatically ensures that the scalar field is not physical. If $\alpha\neq 1$, the resulting theory \eqref{action}, \eqref{einst-eom} and \eqref{phi-eom} is not LS invariant. In this case the current conservation equation: 

\begin{align} \nabla_\mu j^\mu_N=0\;\Rightarrow\;\nabla^2\phi^2=0,\nonumber\end{align} entails that the theory has global scale symmetry instead. Notice that, in the $\alpha\neq 1$ case, the scalar field is a physical DOF, which obeys an specific EOM. Hence, there is not LS symmetry to compensate for the additional DOF associated to $\phi$. The fact that the scalar field is a ghost for $\alpha\neq 1$ represents a real problem for the resulting theory.

%===================================

\subsection{Cosmological example}

%===================================

In order to illustrate the above explanation in a very simple situation we shall consider a cosmological example. Let us substitute the Friedmann-Robertson-Walker metric with flat spatial sections: $ds^2=-dt^2+a^2(t)\delta_{ij}dx^idx^j$ ($i,j=1,2,3$), where $t$ is the cosmic time and $a=a(t)$ is the scale factor, into equations \eqref{einst-eom} and \eqref{phi-eom} with $\alpha=1$. We obtain the following equations:

\bea &&\left(H+\frac{\dot\phi}{\phi}\right)^2=\frac{\lambda}{2}\phi^2,\label{fried}\\
&&2\dot H+3H^2+2\frac{\ddot\phi}{\phi}+4H\frac{\dot\phi}{\phi}-\frac{\dot\phi^2}{\phi^2}=\frac{3\lambda}{2}\phi^2,\label{raycha}\\
&&\frac{\ddot\phi}{\phi}+3H\frac{\dot\phi}{\phi}+\dot H+2H^2=\lambda\phi^2,\label{kg}\eea where $H\equiv\dot a/a$ is the Hubble parameter and the dot means derivative with respect to the cosmic time $t$. It can be shown that the linear combination of equations \eqref{raycha} and \eqref{kg} yields equation \eqref{fried}. Hence, of these three equations only two, say \eqref{fried} and \eqref{kg} are independent. The combination of these equations leads to: $d\ln\xi=d\ln\phi$, where we have introduced the new variable $\xi\equiv H+\dot\phi/\phi$. Integration of this equation yields $H+\dot\phi/\phi=C_0\phi$, where $C_0$ is an integration constant. Comparing the latter equation with \eqref{fried} one gets $C_0=\sqrt{\lambda/2}$. Therefore, only one of the cosmological equations \eqref{fried}, \eqref{raycha} and \eqref{kg}, is an independent EOM. Equation \eqref{fried} can be written in the following manifest LS invariant form: $v'=\pm\sqrt{\lambda/2}v^2,$ where we have introduced the LS invariant variable $v\equiv a\phi$, and the comma means derivative with respect to the Weyl invariant conformal time $\tau=\int dt/a$. Integration of this equation yields:

\bea a(\tau)\phi(\tau)=\mp\frac{\sqrt{\lambda/2}}{\tau-\tau_0},\label{sol}\eea where $\tau_0$ is an integration constant. This is most we can get from the equations of motion. Hence, in order to get an specific solution one must either postulate the functional form of the scalar field $\phi(\tau)$ -- the scale factor is then determined from \eqref{sol} -- or vice versa. The freedom to choose $\phi=\phi(\tau)$ is a consequence of the LS invariance of the equations of motion \eqref{einst-eom} and \eqref{phi-eom}, for $\alpha=1$.

%==========================================================

\subsection{Quadratic model}\label{subsect-r2}

%==========================================================

Let us now consider a second example given by the pure quadratic gravitational model ($\gamma$ is an overall free parameter):

\bea S=\frac{\gamma}{2}\int d^{4}x \sqrt{-g}R^{2}.\label{R2Model}\eea  We can ``linearize'' this action by introducing a new auxiliary field $\phi$, as follows: 

\bea S=\frac{1}{2}\int d^{4}x\sqrt{-g}\left[\frac{\alpha}{6}\phi^2 R-\frac{\lambda}{4}\phi^4\right].\label{recastR2}\eea The EOM for $\phi$ has a non trivial solution given by $\phi^2=\alpha R/3\lambda$ which, when substituted back in \eqref{recastR2}, recovers the original form of \eqref{R2Model} with $\gamma=\alpha^2/36\lambda$. In the ``linear'' form \eqref{recastR2}, the scalar field $\phi$ lacks a kinetic energy density term making it appear to be non dynamical. The following EOM are obtained from action \eqref{recastR2} by varying with respect to the metric: 

\bea \alpha\phi^2G_{\mu\nu}-\alpha\left(\nabla_{\mu}\nabla_{\nu}-g_{\mu\nu}\nabla^2\right)\phi^2+\frac{3\lambda}{4}g_{\mu\nu}\phi^4=0,\label{R2-eom}\eea and the constraint equation

\bea -\alpha R+3\lambda\phi^2=0.\label{R2-const}\eea Combining this last equation with the trace of \eqref{R2-eom}:

\bea -\alpha R+3\alpha\frac{\nabla^2\phi^2}{\phi^2}+3\lambda\phi^2=0,\nonumber\eea one gets the vacuum Klein-Gordon equation:

\bea \alpha\nabla^2\phi^2=0\;\Rightarrow\;\nabla_\mu j_N^\mu=0,\label{kg-eq'}\eea where we have introduced the conserved current $j_N^\mu\equiv 2\alpha\phi\nabla^\mu\phi$. 

Let us notice that neither the action \eqref{R2Model} nor its ``linearization'' \eqref{recastR2} are LS invariant. We have had to clearly state this from the beginning, however, we missed this point on purpose: Even if the action \eqref{recastR2} is not local scale invariant, it is global scale invariant instead. Then one can apply the procedure of Ref. \cite{ferreira-2017} to get the expression for the noether current which, in the present case, is $j_N^\mu\equiv 2\alpha\phi\nabla^\mu\phi$. This current does not vanish unless $\alpha=0$. But this is forbidden since $\alpha=0$ $\Rightarrow\;\phi=0$, i.e., the auxiliary field vanishes, which does not make sense. 

Although the linearized action \eqref{recastR2} initially appears to involve a non-dynamical scalar field due to the absence of a kinetic energy term for $\phi$, this is not true because the scalar field is governed by the current conservation equation \eqref{kg-eq'}, so that it is a physical DOF. If $\phi$ were genuinely non-dynamical, this would imply a deficiency of degrees of freedom in the linearized action \eqref{recastR2}, since the starting action \eqref{R2Model} is quadratic in the curvature scalar.

%%%%%%%%%%%%%%%%%%%%%%%%%%%%%%%%%%%%%%%%%%%%%%%%%%%%%%%%%%%%%%%%%%%%%%%%

\section{Local Scale Invariance and Weyl geometry}\label{sect-weyl-geom}

%%%%%%%%%%%%%%%%%%%%%%%%%%%%%%%%%%%%%%%%%%%%%%%%%%%%%%%%%%%%%%%%%%%%%%%% 

Up to this point in our discussion we have implicitly assumed riemannian background space $V_4$. In this and the next sections we shall consider modifications of Riemann geometry to describe the geometric properties of background spaces. In this section we shall investigate the Weyl geometric (torsionless) modification, while in the next sections, the Riemann-Cartan spaces will be considered. Our aim is to show that the geometric properties of spacetime modify the analysis of LS invariance.

Weyl geometry space $W_4$ is defined as the class of four-dimensional (torsionless) manifolds ${\cal M}_4$ that are paracompact, Hausdorff, connected $C^\infty$, endowed with a locally Lorentzian metric $\bf g$ that obeys the vectorial nonmetricity condition:

\bea \tilde\nabla_\alpha g_{\mu\nu}=-Q_\alpha g_{\mu\nu},\label{vect-nm}\eea where $Q_\alpha$ is the nonmetricity (also Weyl gauge) vector and the covariant derivative $\tilde\nabla_\mu$ is defined with respect to the torsion-free affine connection of the manifold: 

\bea \Gamma^\alpha_{\;\;\mu\nu}=\{^\alpha_{\mu\nu}\}+L^\alpha_{\;\;\mu\nu},\label{gen-aff-c}\eea where 

\bea \{^\alpha_{\mu\nu}\}:=\frac{1}{2}g^{\alpha\lambda}\left(\der_\nu g_{\mu\lambda}+\der_\mu g_{\nu\lambda}-\der_\lambda g_{\mu\nu}\right),\label{lc-aff-c}\eea is the Levi-Civita (LC) connection, while 

\bea L^\alpha_{\;\;\mu\nu}:=\frac{1}{2}\left(Q_\mu\delta^\alpha_\nu+Q_\nu\delta^\alpha_\mu-Q^\alpha g_{\mu\nu}\right),\label{disf-t}\eea is the disformation tensor. The Weyl gauge vector $Q_\alpha$ measures how much the length of given timelike vector varies during parallel transport.

Let us consider the following quite general gravitational action in $W_4$ space \cite{ghile-2022, ghile-2023, harko-2023, quiros-2023-1}:

\begin{align} S_\text{grav}=\frac{1}{2}\int d^4x\sqrt{-g}&\left[\frac{\alpha}{6}\phi^2\tilde R+\omega(\der^*\phi)^2\right.\nonumber\\
&\left.-\frac{\lambda}{4}\phi^4-\frac{\beta^2}{2}Q_{\mu\nu}^2\right],\label{w4-action}\end{align} where $\alpha$, $\omega$, $\lambda$ and $\beta^2$ are free coupling parameters and we have introduced the following notation: $(\der^*\phi)^2\equiv g^{\mu\nu}\der^*_\mu\phi\der^*_\nu\phi$, where $\der^*_\mu\phi=\der_\mu\phi-Q_\mu\phi/2$ is the Weyl gauge derivative of the scalar field, and $Q^2_{\mu\nu}\equiv Q_{\mu\nu}Q^{\mu\nu}$, with the nonmetricity field strength defined as $Q_{\mu\nu}:=2\der_{[\mu}Q_{\nu]}=\der_\mu Q_\nu-\der_\nu Q_\mu$. In \eqref{w4-action} $\tilde R$ is the curvature scalar of $W_4$, which is one of the possible contractions of the curvature tensor of Weyl space:

\bea \tilde R^\alpha_{\;\;\sigma\mu\nu}:=\der_\mu\Gamma^\alpha_{\;\;\nu\sigma}-\der_\nu\Gamma^\alpha_{\;\;\mu\sigma}+\Gamma^\alpha_{\;\;\mu\lambda}\Gamma^\lambda_{\;\;\nu\sigma}-\Gamma^\alpha_{\;\;\nu\lambda}\Gamma^\lambda_{\;\;\mu\sigma}.\nonumber\eea The action \eqref{w4-action} in invariant under the LS transformations \eqref{weyl-t} plus the following gauge transformation of the nonmetricity vector:\footnote{Gauge transformations have also been found to be required when considering LS invariance in Ref. \cite{Sauro-2022}.}

\begin{align} Q_\mu\rightarrow Q_\mu-2\der_\mu\ln\Omega.\label{gauge-t}\end{align} In this section when we speak of LS symmetry we refer to invariance under the Weyl rescalings \eqref{weyl-t} plus the gauge transformations \eqref{gauge-t}.

If write $\tilde R$ in terms of LC quantities:

\bea \tilde R=R-\frac{3}{2}Q_\mu Q^\mu-3\nabla_\mu Q^\mu,\label{ricci-decomp}\eea where $R$ and the covariant derivative $\nabla_\mu$ are given in terms of the LC affine connection \eqref{lc-aff-c}, then the action \eqref{w4-action} can be written in the following equivalent form:

\begin{align} S_\text{grav}&=\frac{1}{2}\int d^4x\sqrt{-g}\left[\frac{\alpha}{6}\phi^2 R+\omega(\der\phi)^2-\frac{\beta^2}{2}Q^2_{\mu\nu}\right.\nonumber\\
&\left.+\frac{\omega-\alpha}{4}\phi^2Q_\mu Q^\mu+\frac{\omega-\alpha}{2}\phi^2\nabla_\mu Q^\mu-\frac{\lambda}{4}\phi^4\right],\label{g-action}\end{align} or, alternatively, as: 

\bea S_\text{grav}=\int d^4x\sqrt{-g}{\cal L}_\text{grav},\;{\cal L}_\text{grav}={\cal L}_\text{ccs}+{\cal L}_\text{stueck},\label{equiv-action}\eea where the conformally coupled scalar and Stueckelberg-type Lagrangians: ${\cal L}_\text{ccs}$ and ${\cal L}_\text{stueck}$, are given by

\begin{align} &{\cal L}_\text{ccs}=\frac{\alpha}{2}\left[\frac{1}{6}\phi^2 R+(\der\phi)^2-\frac{\lambda}{4\alpha}\phi^4\right],\label{ccs-lag}\\
&{\cal L}_\text{stueck}=\frac{\beta^2}{4}\left[-Q^2_{\mu\nu}+\frac{\omega-\alpha}{2\beta^2}\phi^2\left(Q_\mu-\frac{\der_\mu\phi^2}{\phi^2}\right)^2\right],\label{stueck-lag}\end{align} respectively. In the above equations we used the notation $(a_\mu-b_\mu)^2\equiv a_\mu a^\mu-2a_\mu b^\mu+b_\mu b^\mu$. Notice that the Stueckelberg-type Lagrangian ${\cal L}_\text{stueck}$ differs from the standard Stueckelberg Lagrangian in the absence of a gauge fixing term \cite{stueck-1, stueck-2, stueck-3, stueck-4}. Moreover, it is not a typical Proca Lagrangian thanks to the gradient $\der_\mu\phi^2/\phi^2$ within round brackets squared. This leads to the Lagrangian density $\sqrt{-g}{\cal L}_\text{stueck}$ being invariant under \eqref{weyl-t} in contrast to just Proca term which is not LS invariant. As a matter of fact the units $\sqrt{-g}{\cal L}_\text{ccs}$ and $\sqrt{-g}{\cal L}_\text{stueck}$, separately, are both manifestly invariant under the LS transformations \eqref{weyl-t} and \eqref{gauge-t}.

 It is apparent that, unless either $\alpha=\omega$ or $\beta^2\rightarrow\infty$, the effective (point-dependent) square mass of the Proca field: 

\bea m^2_Q(x)=\frac{3(\alpha-\omega)}{2\beta^2}M^2_\text{pl}(x),\nonumber\eea where the point-dependent square Planck mass $M^2_\text{pl}=\phi^2/6$ sets the grand unification scale point by point in spacetime, is a very large quantity $m_Q\sim M_\text{pl}$, so that the vector interactions due to $Q_\mu$, are (effectively) short range interactions with range $\sim M_\text{pl}^{-1}$. This means that the nonmetricity vector field decouples from the low-energy gravitational spectrum. As a consequence, the effective geometry of spacetime is riemannian.

%===================================

\subsection{Equations of Motion}

%===================================

From the action \eqref{g-action} the following EOM can be derived. Vanishing variation of the action with respect to the metric $\delta_{\bf g}S_\text{grav}=0$ leads to the Einstein's EOM:

\begin{align} &\frac{\alpha}{6}\phi^2G_{\mu\nu}-\frac{\alpha}{6}\left(\nabla_\mu\nabla_\nu-g_{\mu\nu}\nabla^2\right)\phi^2\nonumber\\
&+\omega\left[\der_\nu\phi\der_\nu\phi-\frac{1}{2}g_{\mu\nu}(\der\phi)^2\right]\nonumber\\
&+\frac{\omega-\alpha}{4}\phi^2\left(Q_\mu Q_\nu-\frac{1}{2}g_{\mu\nu}Q_\lambda Q^\lambda\right)\nonumber\\
&-\frac{\omega-\alpha}{2}\left[\der_{(\mu}\phi^2Q_{\nu)}-\frac{1}{2}g_{\mu\nu}\der_\lambda\phi^2Q^\lambda\right]\nonumber\\
&-\beta^2\left(Q_\mu^{\;\;\lambda}Q_{\nu\lambda}-\frac{1}{4}g_{\mu\nu}Q^2_{\lambda\sigma}\right)+\frac{\lambda}{8}\phi^4g_{\mu\nu}=0,\label{g-eom}\end{align} where $\der_{(\mu}\phi^2Q_{\nu)}\equiv(\der_\mu\phi^2Q_\nu+\der_\nu\phi^2Q_\mu)/2$ means symmetrization. By requiring vanishing of the variation of the action \eqref{g-action} with respect to the nonmetricity vector: $\delta_{\bf Q}S_\text{grav}=0$, one obtains the following EOM:\footnote{This equation can be written as an inhomogeneous Proca equation for a massive nonmetricity vector field: $$\nabla^\nu Q_{\mu\nu}+m^2_QQ_\mu=j_\mu,$$ where $$m^2_Q=\frac{\alpha-\omega}{4\beta^2}\phi^2,$$ is the effective square mass of the Proca field and $$j_\mu=\frac{\alpha-\omega}{4\beta^2}\der_\mu\phi^2,$$ is a ``Stueckelberg'' current required to preserve the local scale symmetry.}

\begin{align} \nabla^\nu Q_{\mu\nu}=\frac{\omega-\alpha}{4\beta^2}\left(\phi^2Q_\mu-\der_\mu\phi^2\right).\label{proca-eom}\end{align} Finally, vanishing variation of the action \eqref{g-action} with respect to the scalar field $\delta_\phi S_\text{grav}=0$ leads to the following Klein-Gordon (KG) type EOM:

\begin{align} &\frac{\alpha}{6}\phi^2R+\omega(\der\phi)^2-\frac{\omega}{2}\nabla^2\phi^2+\frac{\omega-\alpha}{4}\phi^2Q_\mu Q^\mu\nonumber\\
&+\frac{\omega-\alpha}{2}\phi^2\nabla_\mu Q^\mu-\frac{\lambda}{2}\phi^4=0.\label{kg-eom}\end{align} Comparing this equation with the trace of \eqref{g-eom} one gets that,

\begin{align} (\omega-\alpha)\left[\nabla_\mu(\phi^2Q^\mu)-\nabla^2\phi^2\right]=0,\label{cond}\end{align} which can be written in an equivalent way:

\begin{align} \nabla_\mu j^\mu_N=0,\label{j-cons}\end{align} where we have defined the noether current,

\begin{align} j^\mu_N:=(\omega-\alpha)\left(\phi^2Q^\mu-\nabla^\mu\phi^2\right).\label{j-noether}\end{align} Equation \eqref{cond} can be obtained, alternatively, by taking the divergence of equation \eqref{proca-eom}, if realize that due to anti-symmetry of the field strength tensor: $\nabla_\mu\nabla_\nu Q^{\mu\nu}=0$. 

%---------------------------------------- 

For $\alpha\neq\omega$, the theory \eqref{g-eom}, \eqref{proca-eom} and \eqref{kg-eom}, with nonvanishing noether current $j^\mu_N\neq 0$, is LS invariant, which means that the Weyl symmetry is dynamical. This supports the conclusion of Ref. \cite{oda-2022} that the Weyl transformation is non-dynamical if it does not contain the derivative term of the transformation parameter $\epsilon(x)\equiv\ln\Omega(x)$. According to \cite{oda-2022}, when the LS transformations contain the derivative of the transformation parameter, as in equation \eqref{gauge-t}: $Q_\mu\rightarrow Q_\mu-2\der_\mu\epsilon(x)$, the noether current associated with LS symmetry is nonvanishing, so that the symmetry is dynamical.

%------------------------------------------------------------------------

Notice that when $\alpha\neq\omega$, the current conservation equation \eqref{cond} is not an independent equation, since it is a consequence of the EOM \eqref{proca-eom} and of the identity $\nabla^\mu\nabla^\nu Q_{\mu\nu}=0$. Then, by substituting \eqref{cond} into the trace of Einstein's EOM \eqref{g-eom} one gets \eqref{kg-eom}. This means that the scalar field does not obey and independent EOM, i. e., $\phi$ is a not a physical DOF. In this case equation \eqref{cond}, which can be written in the following convenient way:

\bea \nabla_\mu Q^\mu=\frac{\nabla^2\phi^2}{\phi^2}-\frac{\der_\mu\phi^2}{\phi^2}Q^\mu,\nonumber\eea amounts to a constraint on the nonmetricity vector field. As a consequence, only three of the four components $Q_\mu,$ are independent functions: the two transverse plus the longitudinal polarizations of the (effectively) massive Proca boson. 

As long as local scale invariance is a symmetry of \eqref{equiv-action}, the noether current vanishes in the following cases (we assume that $\beta^2$ is finite): 

\begin{enumerate}

 \item When the coupling constants $\alpha=\omega$, we have that $j^\mu=m^2_Q=0$, which means that $j^\mu_N=0$, so that the conservation equation \eqref{j-cons} becomes into a vanishing identity $0=0$. In this case the nonmetricity field amounts to a massless radiation field propagating in background Riemann space $V_4$. The KG equation \eqref{kg-eom} coincides with the trace of Einstein's equation \eqref{g-eom}, so that the scalar field $\phi$ does not obey an independent EOM, i. e., it is not a physical DOF. Therefore, the number of DOF is the same as in general relativity (GR).

 \item When $\alpha\neq\omega$ but the following equality takes place:
 
 \bea Q_\mu=\frac{\der_\mu\phi^2}{\phi^2}=2\frac{\der_\mu\phi}{\phi},\label{wig-cond}\eea the Weyl space $W_4$ transforms into Weyl integrable geometry (WIG) space (gradient nonmetricity). In this case equation \eqref{cond} amounts to a vanishing identity: $0=0$. The Einstein's EOM \eqref{g-eom} transforms into the following EOM:

 \begin{align} &\phi^2G_{\mu\nu}-\left(\nabla_\mu\nabla_\nu-g_{\mu\nu}\nabla^2\right)\phi^2\nonumber\\
&+6\left[\der_\mu\phi\der_\nu\phi-\frac{1}{2}g_{\mu\nu}(\der\phi)^2\right]+\frac{3\lambda}{4\alpha}\phi^4g_{\mu\nu}=0,\label{g-eom'}\end{align} Then, Eq. \eqref{kg-eom}, which in this case transforms into:

\bea \phi^2R+6(\der\phi)^2-3\nabla^2\phi^2-\frac{3\lambda}{\alpha}\phi^4=0,\label{kg-eom'}\eea coincides with the trace of EOM \eqref{g-eom'}, so that the scalar field does not obey an independent differential equation (it is not a physical DOF). In this case, as in the above item, the number of physical DOF is the same as in GR.
 
\end{enumerate} When $\beta=0$, the action \eqref{g-action} simplifies. Variation of this action with respect to the metric yields EOM \eqref{g-eom} with $\beta^2=0$, while its variation with respect to $Q_\mu$ leads to: $Q_\mu=\der_\mu\phi^2/\phi^2$, so that we recover the case with $\alpha\neq\omega$ described above in item 2.

We may conclude that, in the presence of LS invariance, the noether current vanishes when there are not new physical degrees of freedom in addition to the two polarizations of the massless graviton. However, in the general situation, when we have the theory \eqref{g-eom}, \eqref{proca-eom} and \eqref{kg-eom}, over Weyl space with vectorial nonmetricity $Q_\mu$, there are three DOF in addition to the two polarizations of the massless graviton: the two transversal and a longitudinal polarizations of the massive proca field $Q_\mu$. Therefore, LS invariance requires that the noether current does not vanish, which means that it is a dynamical symmetry. In the backgrounds of riemannian geometry this does not occur, since the gauge transformation of the non-metricity vector \eqref{gauge-t} is necessary for the noether current not to vanish.

%======================================================================

\subsection{The $\tilde R^2$ model in Weyl geometry}
\label{subsect-wg-r2}

%======================================================================

Let us consider the quadratic action \cite{ghile-2022, ghile-2023, harko-2023}:

\bea S=\frac{\gamma}{2}\int d^4x\sqrt{-g}\tilde R^2,\label{wg-r2-action}\eea which operates in $W_4$ space, so that $\tilde R$ is the Weyl-Ricci scalar. It is related with riemannian quantities through Eq. \eqref{ricci-decomp}. If we introduce the auxiliary field $\phi$ as in subsection \ref{subsect-r2}:

\bea S=\frac{1}{2}\int d^4x\sqrt{-g}\left(\frac{\alpha}{6}\phi^2\tilde R-\frac{\lambda}{4}\phi^4\right),\label{wg-lin-action}\eea the quadratic action \eqref{wg-r2-action} results ``linearized.'' Actually, variation of \eqref{wg-lin-action} with respect to the auxiliary field, yields: $\phi^2=\alpha\tilde R/3\lambda$. Hence, if set the latter equality back into \eqref{wg-lin-action}, and set $\gamma=\alpha^2/36\lambda$, the starting action \eqref{wg-r2-action} is recovered. Unlike the $R^2$-model given by the action \eqref{recastR2} over Riemann space $V_4$, the theory \eqref{wg-lin-action} is indeed LS invariant. Notice that \eqref{wg-lin-action} -- with the addition of the term $\propto Q_{\mu\nu}^2$ -- is a particular case of \eqref{w4-action} when the coupling constant $\omega=0$. This case will be discussed in Section \ref{sect-discu}.

%====================================================

%%%%%%%%%%%%%%%%%%%%%%%%%%%%%%%%%%%%%%%%%%%%%%%%%%%%%%%%%%%%%%%%%%%%

\section{Local Scale Invariance and Riemann-Cartan Geometry}
\label{sect-ecg}

%%%%%%%%%%%%%%%%%%%%%%%%%%%%%%%%%%%%%%%%%%%%%%%%%%%%%%%%%%%%%%%%%%%%%

In this and in the next sections we explore LS symmetry within the context of Riemann-Cartan geometry space $U_4$, which is described by a generalized connection \cite{shapiro-rev, obukhov-pla-1982, german-1985, rodrigo-2023}: 

\bea \Gamma^\alpha_{\;\;\mu\nu}=\{^\alpha_{\mu\nu}\}+K^\alpha_{\;\;\mu\nu},\label{gen-aff-c'}\eea where $\{^\alpha_{\mu\nu}\}$ is the LC connection, while the contortion tensor is defined as 

\bea K_{\alpha\beta\gamma}=\frac{1}{2}\left(T_{\alpha\beta\gamma}+T_{\gamma\alpha\beta}+T_{\beta\alpha\gamma}\right).\label{contort-t}\eea  The torsion tensor is defined as the antisymmetric part of the connection

\bea T^{\lambda}_{\mu\nu}:=2\Gamma^{\lambda}_{\;\;[\mu\nu]}=\Gamma^{\lambda}_{\ \mu\nu}-\Gamma^{\lambda}_{\ \nu\mu}.\label{torsion}\eea 

In terms of Cartan variables: the tetrad fields $e^\mu_a$ and the spin connection $\omega^\alpha_{\;\;a\mu}=\der_\mu e^\alpha_a+\Gamma^\alpha_{\;\;\lambda\mu}e^\lambda_a$, the torsion tensor can be written as:\footnote{Greek letters $\alpha$, $\beta$,...,$\mu$,... etc., stand for spacetime indices, while small latin letters $a$, $b$, $c$,... etc., represent the tangent space or internal indices.}

\bea T^\alpha_{\;\;\mu\nu}=\omega^\alpha_{\;\;[\mu\nu]}+e^\alpha_a\der_{[\mu}e^a_{\nu]}.\label{torsion-tetrad}\eea In the similar way the generalized curvature tensor or, just curvature tensor of the connection;\footnote{In what follows quantities and operators with the tilde are defined with respect to the Riemann-Cartan connection \eqref{gen-aff-c}. Meanwhile, the Riemann-Christoffel curvature tensor $R^\alpha_{\;\;\mu\beta\nu}$ and other riemannian quantities like the covariant derivative operator $\nabla_\mu$, represent torsionless quantities and operators which are defined with respect to the LC connection. Notice that this notation is not applied to the torsion tensor itself and the related quantities.} 

\bea \tilde R^\alpha_{\;\;\mu\beta\nu}=\der_\beta\Gamma^\alpha_{\;\;\mu\nu}-\der_\nu\Gamma^\alpha_{\;\;\mu\beta}+\Gamma^\alpha_{\;\;\sigma\beta}\Gamma^\sigma_{\;\;\mu\nu}-\Gamma^\alpha_{\;\;\sigma\nu}\Gamma^\sigma_{\;\;\mu\beta},\label{curv-t}\eea can be written in terms of Cartan variables. In particular, the curvature tensor with two internal indices and two spacetime indices reads:

\bea \tilde R^a_{\;\;b\mu\nu}=\der_\mu\omega^a_{\;\;b\nu}-\der_\nu\omega^a_{\;\;b\mu}+\omega^a_{\;\;c\mu}\omega^c_{\;\;b\nu}-\omega^a_{\;\;c\nu}\omega^c_{\;\;b\mu}.\label{curv-t-tetrad}\eea 

The curvature tensor \eqref{curv-t} can be decomposed into the LC (riemannian) and contortion contributions as \cite{langvik-prd-2021}:

\bea &&\tilde R^\alpha_{\;\;\mu\beta\nu}=R^\alpha_{\;\;\mu\beta\nu}+\nabla_\beta K^\alpha_{\;\;\nu\mu}-\nabla_\nu K^\alpha_{\;\;\beta\mu}\nonumber\\
&&\;\;\;\;\;\;\;\;\;\;\;\;\;\;\;+K^\alpha_{\;\;\beta\lambda}K^\lambda_{\;\;\nu\mu}-K^\alpha_{\;\;\nu\lambda}K^\lambda_{\;\;\beta\mu}.\label{curv-t-decomp}\eea Two geometrical scalars can be obtained from the curvature tensor \eqref{curv-t} by contraction of all of its indices: 

1) The generalized Ricci scalar

\bea \tilde R\equiv g^{\alpha\lambda}g^{\mu\nu}\tilde R_{\alpha\mu\lambda\nu}=R+{\cal T}+2\nabla_\mu T^\mu,\label{curv-scalar}\eea where 

\bea {\cal T}=\frac{1}{4}T_{\lambda\mu\nu}T^{\lambda\mu\nu}-\frac{1}{2}T_{\lambda\mu\nu}T^{\nu\lambda\mu}-T_\mu T^\mu,\label{t-scalar}\eea is the torsion scalar, while $T^\mu\equiv g_{\lambda\nu}T^{\lambda\mu\nu}$ is the torsion vector, and 

2) The Holst term \cite{Hojman-1980, holst-1996}:

\bea {\cal R}\equiv\frac{1}{2}e^{\alpha\mu\lambda\nu}\tilde R_{\alpha\mu\lambda\nu}=-3\nabla_\mu\hat T^\mu+\frac{1}{4}e^{\alpha\beta\mu\nu}T_{\lambda\alpha\beta}T^{\lambda}_{\ \mu\nu},\label{holst}\eea where $e^{\alpha\mu\lambda\nu}=\epsilon^{\alpha\mu\lambda\nu}/\sqrt{-g}$ is the totally antisymmetric unit pseudo-tensor and $\epsilon^{\alpha\mu\lambda\nu}$ is the totally antisymmetric LC permutation symbol ($\sqrt{-g}e^{0123}=+1$), while $\hat T^\alpha\equiv e^{\alpha\mu\nu\lambda}T_{\mu\nu\lambda}/6$ is known as axial vector. Equation \eqref{holst} is a rewriting of the Nieh-Yan identity in Riemann-Cartan geometry \cite{nieh-yan-1982, nieh-yan-2007, banerjee-2010}:

\bea &&\der_\mu\left(\sqrt{-g}e^{\mu\nu\lambda\sigma}T_{\nu\lambda\sigma}\right)=6\sqrt{-g}\nabla_\mu\hat T^\mu=\nonumber\\
&&\sqrt{-g}\left(\frac{1}{2}e^{\mu\nu\lambda\sigma}T^\rho_{\;\;\mu\lambda}T_{\rho\nu\sigma}-e^{\mu\nu\lambda\sigma}\tilde R_{\mu\nu\lambda\sigma}\right).\label{nieh-yan-id}\eea

%======================================================

\subsection{Local scale transformations of Torsion}

%=====================================================

Under \eqref{weyl-t} the tetrad vectors transform like:

\bea e_\mu^a\rightarrow\Omega(x) e_\mu^a\;\left(e^\mu_a\rightarrow\Omega^{-1}(x)e^\mu_a\right),\;e\rightarrow\Omega^4(x)e,\label{conf-t-tetrad}\eea where $e=\text{det}|e_\mu^a|=\sqrt{-g}$. Hence, in order to determine the transformation law for the torsion tensor and for its contractions, let us consider the definition \eqref{torsion-tetrad}. There are two possible different scenarios \cite{nieh-yan-2007}:

\begin{enumerate}

\item Both the spin connection $\omega^a_{\;\;b\mu}$ and the tetrad $e^\mu_a$ are taken to be independent field variables and are to be determined by the theory. In this case, in addition to the transformation of the tetrad \eqref{conf-t-tetrad}, one is free to postulate the transformation properties of the spin connection with respect to the conformal transformations \eqref{weyl-t}. Here we assume that under \eqref{weyl-t} the spin connection transforms in the following way:

\bea \omega^\alpha_{\;\;\mu\nu}\rightarrow\omega^\alpha_{\;\;\mu\nu}+(2s-1)\delta^\alpha_\nu\der_\mu\ln\Omega-g_{\mu\nu}\der^\alpha\ln\Omega,\label{conf-t-w1}\eea where $s$ is a real free parameter. This leads to the following transformation law of the torsion tensor \eqref{torsion-tetrad} under the conformal transformations:\footnote{A similar transformation of the torsion is considered in \cite{albanese}, where the conformal equivalence between a minimally coupled and a non-minimally coupled scalar field theories is studied in $U_4$ space. In Ref. \cite{zhytnikov} three types of conformal invariance in $d$-dimensional manifolds with curvature, nonmetricity and torsion at once, are discussed: 1) when the tensorial part of the connection is not transformed (type I), 2) when the nonmetricity is not transformed but only the trace of the torsion $T_\mu=T^\lambda_{\;\;\mu\lambda}$ is modified by the conformal transformation (type II) and 3) when the connection itself is not transformed. The transformation law \eqref{conf-t-torsion} with $s\neq 1$ does not belong in none of these types, however, when $s=1$ it belongs in type III transformation in the classification of \cite{zhytnikov}.}

\bea T^\alpha_{\;\;\mu\nu}\rightarrow T^\alpha_{\;\;\mu\nu}+2s\delta^\alpha_{[\nu}\der_{\mu]}\ln\Omega.\label{conf-t-torsion}\eea In Ref. \cite{shapiro-rev} the above transformation property of the torsion tensor is called as ``strong conformal symmetry.'' Meanwhile, the case when the torsion is not transformed by the conformal transformations \eqref{weyl-t}, which corresponds to the choice $s=0$ in \eqref{conf-t-torsion}, is called in that reference as ``weak conformal symmetry.''

\item The spin connection are given through the Ricci coefficients of rotation in terms of the tetrad:

\bea \omega_{ab\mu}=\frac{1}{2}e^c_\mu\left(\gamma_{cab}-\gamma_{abc}-\gamma_{bca}\right),\label{w2}\eea where, following the notation of \cite{nieh-yan-2007}, we have defined:

\bea \gamma^a_{\;\;bc}:=\left(e^\mu_be^\nu_c-e^\mu_ce^\nu_b\right)\der_\nu e^c_\mu.\label{def-w2}\eea In this case, taking into account the conformal transformation of the tetrad \eqref{conf-t-tetrad}, we have that under \eqref{weyl-t} the spin connection transform like:

\bea \omega^\alpha_{\;\;\mu\nu}\rightarrow\omega^\alpha_{\;\;\mu\nu}+\delta^\alpha_\nu\der_\mu\ln\Omega-g_{\mu\nu}\der^\alpha\ln\Omega,\label{conf-t-w2}\eea where $\omega^\alpha_{\;\;\mu\nu}=e^{a\alpha}e^b_\mu\omega_{ab\nu}$. Unfortunately, in this case the affine connection of the manifold: $\Gamma^\alpha_{\;\;\mu\nu}=\left(\der_\nu e^a_\mu+\omega^a_{\;\;b\nu}e^b_\mu\right)e^\alpha_a=\{^\alpha_{\mu\nu}\}$, coincides with the LC connection so that the torsion tensor vanishes. Therefore, the second scenario is not of help in defining the transformation properties of the torsion under the conformal transformations \eqref{weyl-t}.

\end{enumerate} 

In what follows we shall consider that the torsion tensor transforms according to the law \eqref{conf-t-torsion} under the LS transformations \eqref{weyl-t}. In consequence, under these transformations, the torsion vector and the axial vector transform like:

\bea &&T_\mu\rightarrow T_\mu+3s\der_\mu\ln\Omega,\nonumber\\
&&\hat T_\mu\rightarrow\hat T_\mu,\label{conf-t-tt}\eea respectively. Besides, under \eqref{weyl-t} other relevant quantities transform in the following way:

\bea &&{\cal T}\rightarrow\Omega^{-2}\left[{\cal T}-4sT^\mu\der_\mu\ln\Omega-6s^2(\der\ln\Omega)^2\right],\label{conf-t-tscalar}\\
&&\nabla_\mu T^\mu\rightarrow\Omega^{-2}\left[\nabla_\mu T^\mu+2T^\mu\der_\mu\ln\Omega\right.\nonumber\\
&&\left.\;\;\;\;\;\;\;\;\;\;\;\;\;\;\;\;\;\;\;\;\;\;\;+6s(\der\ln\Omega)^2+3s\nabla^2\ln\Omega\right],\label{conf-t-div}\eea where ${\cal T}$ is the torsion scalar in \eqref{curv-scalar}. It can be shown that the generalized curvature scalar $\tilde R$, which is given by \eqref{curv-scalar}, transforms as:

\begin{align} \tilde R\rightarrow &\Omega^{-2}\left[\tilde R-4(s-1)T^\mu\der_\mu\ln\Omega\right.\nonumber\\
&\left.-6(s-1)^2(\der\ln\Omega)^2+6(s-1)\nabla^2\ln\Omega\right].\label{conf-t-gcurvs}\end{align} Besides, since under \eqref{weyl-t},

\begin{align} &e^{\alpha\beta\mu\nu}T_{\lambda\alpha\beta}T^\lambda_{\;\;\mu\nu}\rightarrow\Omega^{-2}\left[e^{\alpha\beta\mu\nu}T_{\lambda\alpha\beta}T^\lambda_{\;\;\mu\nu}+24s\hat T^\mu\der_\mu\ln\Omega\right],\nonumber\\
&\nabla_\mu\hat T^\mu\rightarrow\Omega^{-2}\left(\nabla_\mu\hat T^\mu+2\hat T^\mu\der_\mu\ln\Omega\right),\nonumber\end{align} then the Holst term transforms like:

\bea {\cal R}\rightarrow\Omega^{-2}\left[{\cal R}+6(s-1)\hat T^\mu\der_\mu\ln\Omega\right].\label{conf-t-holst}\eea

From equations \eqref{conf-t-gcurvs} and \eqref{conf-t-holst} it follows that the particular case with $s=1$, is outstanding. Actually, for $s=1$ the generalized curvature scalar and the Holst term, both transform in a very simple way: $\tilde R\rightarrow\Omega^{-2}\tilde R$, ${\cal R}\rightarrow\Omega^{-2}{\cal R}$.

%%%%%%%%%%%%%%%%%%%%%%%%%%%%%%%%%%%%%%%%%%%%%%%%%%%%%%%%%%%%%

\section{Scale invariance in Einstein-Cartan-Holst theory}
\label{sect-ecartan}

%%%%%%%%%%%%%%%%%%%%%%%%%%%%%%%%%%%%%%%%%%%%%%%%%%%%%%%%%%%%%

Scale invariance could be a symmetry of the gravitational laws at high energies. Then it makes sense to think of quadratic $U_4$ curvature terms like in the following action:

\bea S=\frac{1}{2\lambda}\int d^4x\sqrt{-g}{\cal P}^2,\label{he-action}\eea where we defined the scalar: ${\cal P}:=\tilde R+\alpha{\cal R}$, with $\lambda$ being a free positive constant. The action \eqref{he-action} is invariant with respect to the LS transformations \eqref{weyl-t} and \eqref{conf-t-torsion} with $s=1$, since ${\cal P}\rightarrow\Omega^{-2}{\cal P}$. We can ``linearize'' the above action through introducing an auxiliary field $\phi$ \cite{wands-1994, tourrenc-1983}, so that the dynamically equivalent action reads:

\bea S=\frac{1}{2}\int d^4x\sqrt{-g}\left(\phi^2{\cal P}-\frac{\lambda}{4}\phi^4\right).\label{x-action}\eea This action inherits the LS symmetry of \eqref{he-action}, due to invariance under \eqref{weyl-t} and \eqref{conf-t-torsion} with $s=1$. It corresponds to a $\omega_{BD}=0$ Brans-Dicke (BD) type model with the inclusion of torsion. We shall call the theory which is based in the action \eqref{x-action}, as Scalar-Einstein-Cartan-Holst (SECH) theory of gravity.

%=================================

\subsection{Boundary terms}

%=================================

The Einstein-Cartan-Holst (ECH) theory of gravity is given by the following generalization of the Einstein-Hilbert (EH) action (here we use units where $8\pi G=M^{-2}_\text{pl}=1$):

\bea S_\text{ECH}=\frac{1}{2}\int d^4x\sqrt{-g}\left(\tilde R+\alpha{\cal R}\right),\label{ec-action}\eea where $\alpha$ is a dimensionless coupling parameter while the Riemann-Cartan curvature scalar $\tilde R$ and the Holst term ${\cal R}$ are given by equations \eqref{curv-scalar} and \eqref{holst}, respectively. In order to derive the equations of motion (EOM) of this theory one should perform first variations of the action \eqref{ec-action} with respect to the independent variables: the metric and the torsion, instead of the tetrads and of the spin connection. Let us consider the ECH Lagrangian:

\bea \tilde L_\text{ECH}=\frac{\sqrt{-g}}{2}\left(\tilde R+\alpha{\cal R}\right).\label{ec-lag}\eea Omitting the total derivatives in \eqref{ec-lag} we can write this Lagrangian in the dynamically equivalent form:

\bea L_\text{ECH}=\frac{\sqrt{-g}}{2}\left(R+{\cal T}+\frac{\alpha}{4}e^{\alpha\beta\mu\nu}T_{\lambda\alpha\beta}T^\lambda_{\;\;\mu\nu}\right).\label{ec-lag-x}\eea The following relationship can be established between the above Lagrangians:

\bea \tilde L_\text{ECH}=L_\text{ech}-\frac{1}{2}\der_\mu\left(\sqrt{-g}\Delta T^\mu\right),\label{rel-lags}\eea where we have defined $\Delta T^\mu:=3\alpha\hat T^\mu-2T^\mu$.

Under the conformal transformations of the metric \eqref{weyl-t} these Lagrangians, however, transform in different ways:

\begin{align} \tilde L_\text{ECH}\rightarrow &\Omega^2\left\{\tilde L_\text{ECH}-(s-1)\sqrt{-g}\left[3(s-1)(\der\ln\Omega)^2\right.\right.\nonumber\\
&\left.\left.-3\nabla^2\ln\Omega-\Delta T^\mu\der_\mu\ln\Omega\right]\right\},\label{conf-t-ec-lag}\end{align} while 

\begin{align} L_\text{ECH}\rightarrow &\Omega^2\left\{L_\text{ECH}+\sqrt{-g}\left[3(1-s^2)(\der\ln\Omega)^2\right.\right.\nonumber\\
&\left.\left.+s\Delta T^\mu\der_\mu\ln\Omega\right]\right\}.\label{conf-t-ec-lag-x}\end{align}

The following choice of the free parameter: $s=1$, is special. In this case, under the conformal transformations \eqref{weyl-t}:

\begin{align} &\tilde L_\text{ECH}\rightarrow\Omega^2\tilde L_\text{ECH},\nonumber\\
&L_\text{ECH}\rightarrow\Omega^2\left(L_\text{ECH}+\sqrt{-g}\Delta T^\mu\der_\mu\ln\Omega\right).\label{transf-lags}\end{align} Equation \eqref{transf-lags} shows that, even in this outstanding case when $s=1$, the dynamically equivalent Lagrangians $\tilde L_\text{ECH}$ and $L_\text{ECH}$ transform in different ways under the conformal transformations. 

The above discrepancy can be explained in the following way. Omitting the divergence term in the RHS of \eqref{rel-lags} amounts to vanishing of the following boundary term:\footnote{We want to mention that, for any fields $\Psi_A$ which vanish on the boundary the integral \eqref{boundary-t} vanishes as well. However, for the torsion and its components (these include $\Delta T^\mu$), the former requirement may not be satisfied, so that vanishing of \eqref{boundary-t} may not be true. In this case we may not drop the divergence in \eqref{rel-lags}, so that the Lagrangians $\tilde L_\text{ech}$ and $L_\text{ech}$ are not dynamically equivalent.} 

\bea -\int_{\der{\cal M}} d^3x\sqrt{-h}n_\mu\Delta T^\mu,\label{boundary-t}\eea in the corresponding action $S_\text{ECH}=\int d^4x\tilde L_\text{ECH}$ over $U_4$ space. In the above expression $h_{\mu\nu}=g_{\mu\nu}+n_\mu n_\nu$ is the metric induced on the boundary $\der{\cal M}$ by the gravitational metric $g_{\mu\nu}$, while $n^\mu$ is a unit vector normal to the boundary at each point. It may happen that, if in the action we take into account the York-Gibbons-Hawking boundary term \cite{york-1972, gib-haw-1977, wands-1994}:

\bea 2\int_{\der{\cal M}} d^3x\sqrt{-h}\tilde K,\label{ygh-boundary-t}\eea where $\tilde K=h^{\mu\nu}K_{\mu\nu}$ with $\tilde K_{\mu\nu}$--the components of the extrinsic curvature of the boundary of ${\cal M}\in U_4$, then it might happen that some of the terms in \eqref{ygh-boundary-t} and the boundary term \eqref{boundary-t} cancel out so that the Lagrangians $\tilde L_\text{ECH}$ and $L_\text{ECH}$ are indeed dynamically equivalent.

In Ref. \cite{casad-ijmpd-2002} a method is developed in order to clarify the interplay between boundary terms and conformal transformations in scalar-tensor theories of gravity, through considering fifth-dimensional equations and Kaluza-Klein compactification. Perhaps a similar procedure could be fruitful in the present case in order to obtain all of the boundary terms. However, this subject is beyond the scope of the present paper.

%%%%%%%%%%%%%%%%%%%%%%%%%%%%%%%%%%%%%%%%%%%%%%%%%%%%%%%%%%%%%%%%%%

\section{Scale invariant SECH theory}\label{sect-ginv-ech}

%%%%%%%%%%%%%%%%%%%%%%%%%%%%%%%%%%%%%%%%%%%%%%%%%%%%%%%%%%%%%%%%%%

In this section we shall assume that the above considerations about boundary terms is correct, so that, even in the presence of torsion, these may be omitted. This approach is followed in several bibliographic references on this subject as, for instance, in Ref. \cite{banerjee-2010}. Here we consider the action \eqref{x-action} or, explicitly:\footnote{The most general metric-scalar-torsion theories of gravity in $n$ dimensions, which are invariant under LS transformation, where investigated in \cite{helayel-neto}. However, the authors considered the particular case when only the trace of the torsion is transformed by the conformal transformation.}

\bea S=\frac{1}{2}\int d^4x\sqrt{-g}\left[\phi^2\left(\tilde R+\alpha{\cal R}\right)-\frac{\lambda}{4}\phi^4\right].\label{sech-action}\eea Notice that, in the present case, boundary terms like \eqref{boundary-t} are replaced by: $-\int_{\der{\cal M}} d^3x\sqrt{-h}\,n_\mu\phi^2\Delta T^\mu,$ which, thanks to the nonminimal coupling of the scalar field with the curvature, vanishes. Hence, there are no problems with vanishing of boundary terms which contain torsion.

The SECH theory resulting from \eqref{sech-action}, is invariant under the LS transformations \eqref{weyl-t} and \eqref{conf-t-torsion} with $s=1$. It is worth to split \eqref{sech-action} into its LC and torsion parts: 

\begin{align} &S=\frac{1}{2}\int d^4x\sqrt{-g}\left[\phi^2\left(R+{\cal T}+\alpha\mathcal{R}\right)\right.\nonumber\\
&\;\;\;\;\;\;\;\;\;\;\;\;\;\;\;\;\;\;\;\;\;\;\;\;\;\;\;\;\;\left.+2\phi^2\nabla_\mu T^\mu-\frac{\lambda}{4}\phi^4\right].\label{splitedx-action}\end{align}

Vanishing variation of the action \eqref{splitedx-action} with respect to the metric $\delta_{\bf g}S=0$, leads to the Einstein-Cartan's EOM (for details of this derivation see appendix \ref{app-a}):

\begin{align} &\phi^{2}G_{\mu\nu}+\phi^2\Big({\cal T}_{\mu\nu}-\frac{1}{2}g_{\mu\nu}{\cal T}\Big)-\alpha\phi^2{\cal R}_{\mu\nu}\nonumber\\
&-\left(\nabla_\mu\nabla_\nu-g_{\mu\nu}\nabla^2\right)\phi^{2}+\frac{\lambda}{8}\phi^4g_{\mu\nu}\nonumber\\
&-2\der_{(\mu}\phi^2T_{\nu)}+g_{\mu\nu}\der_\lambda\phi^2T^\lambda=0,\label{ecart-eom}\end{align} where $G_{\mu\nu}=R_{\mu\nu}-g_{\mu\nu}R/2$ is the Einstein's tensor, $X_{(\mu}Y_{\nu)}\equiv(X_\mu Y_\nu+X_\nu Y_\mu)/2$ means index symmetrization and the quantities ${\cal T}_{\mu\nu}$ and ${\cal R}_{\mu\nu}$ have been defined in equations \eqref{app-a-3} and \eqref{app-a-6} of the appendix \ref{app-a}. Their traces coincide with the torsion scalar ${\cal T}$ and with the Holst term ${\cal R}$, respectively.

Similarly, vanishing variation of \eqref{splitedx-action} with respect to the scalar field $\delta_{\bf \phi}S=0$, and to the connection $\delta_{\Gamma}S=0$, yields the corresponding EOM:

\bea R+{\cal T}+\alpha{\cal R}+2\nabla_\mu T^\mu-\frac{\lambda}{2}\phi^2=0,\label{kg-sech}\eea and (see Ref. \cite{Langvik-2021},)

\begin{align} &T_{\alpha\mu\nu}+2g_{\alpha[\mu}T_{\nu]}+\frac{2}{\phi^2}g_{\alpha[\mu}\der_{\nu]}\phi^2\nonumber\\
&+\alpha\left(e^{\;\;\;\;\;\;\lambda}_{\alpha\mu\nu}T_\lambda+\frac{1}{2}e^{\;\;\;\;\sigma\lambda}_{\alpha\mu}T_{\nu\sigma\lambda}+e^{\;\;\;\;\;\;\lambda}_{\alpha\mu\nu}\frac{\der_\lambda\phi^2}{\phi^2}\right)=0,\label{torsion-eom}\end{align} respectively. Multiplying this equation by $g^{\alpha\mu}$  one obtains:

\begin{align}T_{\mu}=-\frac{3}{2}\frac{\partial_{\mu}\phi^2}{\phi^2},\label{t-vect}\end{align} while, if multiply \eqref{torsion-eom} by $g^{\mu\nu}$ one gets: $\hat{T}_{\mu}=0.$ It may be verified that the following is solution of \eqref{torsion-eom} \cite{Langvik-2021}:

\begin{align} T_{\alpha\mu\nu}=\frac{1}{\phi^2}g_{\alpha[\mu}\der_{\nu]}\phi^2.\label{tor-sol}\end{align} This means that the torsion is dynamical if the scalar field is a dynamical DOF. Hence, the torsion is dynamical if we renounce to LS symmetry. Otherwise, if the SECH theory is LS invariant, the scalar field is not a physical DOF, so that the torsion is not dynamical.

If we compare the trace of \eqref{ecart-eom}

\begin{align} &R+\mathcal{T}+\alpha\mathcal{R}-3\frac{\nabla^2\phi^2}{\phi^2}-\frac{\lambda}{2}\phi^2-2\frac{\der_\lambda\phi^2}{\phi^2}T^\lambda=0,\label{trace-ecart-eom}\end{align} with equation \eqref{kg-sech}, one gets the following noether current conservation equation:

\begin{align} \nabla_\mu j^\mu_N=0,\label{t-noether}\end{align} where we have defined the noether current as: 

\begin{align} j^\mu_N\equiv\phi^2T^\mu+3\der^\mu\phi^2/2.\label{t-noether-c}\end{align} Now, if substitute $T^\mu$ from \eqref{t-vect} into \eqref{t-noether-c} one gets a vanishing noether current $j^\mu_N=0$, so that \eqref{t-noether} becomes an identity $0=0$.\footnote{Notice also that, from equation \eqref{t-vect}, when matching the resulting LS transformation of the torsion \eqref{conf-t-tt} and the scalar field \eqref{weyl-t}, it follows that, necessarily $s=1$.} The present case shares similitude with the CCS theory given by the action \eqref{action} over Riemann space $V_4$ (section \ref{sect-scale-inv}), and also with the theory given by action \eqref{w4-action} over Weyl space $W_4$, for the following choice of the free parameters (section \ref{sect-weyl-geom}): $\alpha=\omega$. In both cases there are not new degrees of freedom in addition to the two polarizations of the massless graviton. As already explained, this is so thanks to the fact that the scalar field is not a physical degree of freedom due, precisely, to invariance under the local scale transformations \eqref{weyl-t}. In the present case, since the torsion obeys an algebraic constraint \eqref{torsion-eom}, it does not carry physical degrees of freedom. Meanwhile, the scalar field does not obey an independent EOM. Actually, if in equation \eqref{kg-sech} -- which is obtained by varying the action with respect to the scalar field -- we take into account \eqref{t-vect}, the resulting equation coincides with the trace of the Einstein's EOM \eqref{ecart-eom}. We can say that the scalar field compensates the gauge freedom due to Weyl invariance.

%-------------------------------------------------

%%%%%%%%%%%%%%%%%%%%%%%%%%%%%%%%%%%%%%%%%%%%%%%%%%%%%%%%%%%%%%%%%%%

\section{CCS theory in Riemann-Cartan and in Weyl-Cartan spaces}
\label{sect-u4-y4}

%%%%%%%%%%%%%%%%%%%%%%%%%%%%%%%%%%%%%%%%%%%%%%%%%%%%%%%%%%%%%%%%%%%

The action \eqref{splitedx-action}, which is associated with Riemann-Cartan space $U_4$, may be further generalized by the inclusion of a kinetic term for the scalar field $\phi$. The resulting action reads:\footnote{In Ref. \cite{hohmann-1, hohmann-2, hohmann-3} the most general classes of teleparallel scalar-torsion theories of gravity where investigated. It was shown that the different theories in these classes are related to each other by conformal transformations of the tetrad and appropriate redefinitions of the scalar field.}

\begin{align} &S=\int d^4x\sqrt{-g}\Big[a\phi^2\left(\Tilde{R}+\alpha\mathcal{R}\right)\nonumber\\
&\;\;\;\;\;\;\;\;\;\;\;\;\;\;\;\;\;\;\;\;\;\;\;\;\;\;\;\;\;+\frac{1}{2}(\der\phi)^2-\frac{\lambda}{4}\phi^4\Big],\label{ccs-t}\end{align} which can be thought as the torsionfull version of \eqref{action} if set the constant $a=1/12$. Regrettably, the addition of a kinetic term for $\phi$ spoils the LS invariance. Recalling that, under \eqref{weyl-t}, the kinetic term transforms as:

\begin{align} (\der\phi)^2\rightarrow\Omega^{-4}\left[(\der\phi)^2-2\phi\der_\mu\phi\der^\mu\ln\Omega+\phi^2(\der\ln\Omega)^2\right],\label{*}\end{align} one can immediately notice that this transformation generates derivative terms of the conformal factor, that are usually cancelled out with similar ones coming from the transformation of the LC Ricci scalar (see equation \eqref{usef-2}.) However, in the presence of torsion, transforming under \eqref{weyl-t} according to \eqref{conf-t-torsion} with $s=1$, the mentioned derivatives coming from the transformation of the riemannian curvature scalar are cancelled out by the ones coming from the transformation of the torsion. Therefore, the derivatives of the conformal factor in \eqref{*} persist in the final result, thus breaking LS invariance.

One way in which we may consider generalization of the models explored so far in this paper, is by replacing the Riemann-Cartan geometry by Weyl-Cartan geometry space, where the connection reads:

\begin{align} \Gamma^\alpha_{\;\;\mu\nu}=\{^\alpha_{\mu\nu}\}+L^\alpha_{\;\;\mu\nu}+K^\alpha_{\;\;\mu\nu}.\label{y4-connect}\end{align} The generalized curvature tensor $\tilde R^\alpha_{\;\;\mu\sigma\nu}$ is now defined with respect to the connection \eqref{y4-connect}. The following action:

\begin{align} &S_g=\int d^4x\sqrt{-g}\left[\frac{1}{6}\phi^2\left(\tilde R+\alpha{\cal R}\right)\right.\nonumber\\
&\left.\;\;\;\;\;\;\;\;\;\;\;\;\;\;\;\;\;\;\;\;\;\;\;\;\;\;\;\;\;\;+\omega(\der^*\phi)^2-\frac{\lambda}{4}\phi^4-\frac{\beta^2}{2}Q_{\mu\nu}^2\right],\label{y4-action}\end{align} where $\der^*_\mu\phi=\der_\mu\phi-Q_\mu\phi/2$ and $\tilde R$ is defined with respect to the connection \eqref{y4-connect}, is invariant under the Weyl scale transformations \eqref{weyl-t} and \eqref{gauge-t}, if the torsion tensor is not transformed. This corresponds to the choice $s=0$ in \eqref{conf-t-torsion}, which is called as weak conformal symmetry in Ref. \cite{shapiro-rev}.

%%%%%%%%%%%%%%%%%%%%%%%%%%%%%%%%%%%%%%%%%%%%%%%%%%%%%%%%

\section{Discussion}\label{sect-discu}

%%%%%%%%%%%%%%%%%%%%%%%%%%%%%%%%%%%%%%%%%%%%%%%%%%%%%%%%

The major achievement of the present paper is to generalize previously obtained results on LS invariance, within the framework of gravitational theories over Riemann space \cite{jackiw-2015, garay, oda-2022}, to theories where the background space has non-riemannian structure; namely Weyl geometric or Riemann-Cartan structure. In Ref. \cite{jackiw-2015} it was shown that LS invariance of CCS theory in $V_4$ background space does not have any associated noether current, so that it is a non-dynamical symmetry. This result was confirmed in \cite{garay} for WTDiff theories over $V_4$ space and then generalized in \cite{oda-2022} to any LS invariant theory of gravity in four-dimensional Riemann space. In the present investigation, on the basis of the computation of the conserved noether current, we have established that, in those LS invariant theories of gravity in $W_4$ and $U_4$ spaces, where the number of gravitational DOF-s is the same as in GR, LS symmetry plays no dynamical role.

In Section \ref{sect-scale-inv} we have reviewed the results of \cite{jackiw-2015, oda-2022} for gravitational theories in Riemann space. In this case the presence of a quadratic term $R^2$ is associated with an additional degree of freedom with respect to GR. Since the theory is not LS invariant, the additional degree of freedom is not compensated by any symmetry. Therefore, the nonvanishing noether current entails that the theory is global scale invariant.

The simplest modification of Riemann geometry is Weyl geometry. In the latter case not only the direction of vectors change during parallel transport but, also, the length of vectors varies from point to point in spacetime. In Section \ref{sect-weyl-geom} we have replaced Riemann background space by Weyl geometry space: $V_4\rightarrow W_4$. In addition, a slight modification of the gravitational action specialized to Weyl geometry structure of background space, is required. The resulting action \eqref{w4-action} represents a large class of theories in which belong theories investigated in \cite{ghile-2022, ghile-2023, harko-2023, quiros-2023-1, quiros-arxiv-23}. Here, in addition to the Weyl transformations \eqref{weyl-t}, a gauge transformation \eqref{gauge-t} of the nonmetricity vector is required. Since the latter transformation includes the derivative of the transformation parameter $\epsilon(x)\equiv\ln\Omega^2(x)$, the noether current does not vanish. This confirms the conclusion in \cite{oda-2022} that vanishing of the noether current is associated with the absence of the derivative $\der_\mu\epsilon(x)$ in the LS transformations, thus generalizing the result to non-riemannian space. Only for specific values of the free parameters $\alpha$ and $\omega$ in \eqref{w4-action}, where the vectorial nonmetricity can be omitted, the resulting local scale invariant theory is accompanied by vanishing noether symmetry. It is also confirmed that, within the framework of LS invariant theories of gravity, vanishing of the noether current means that there are not new gravitational DOF in addition to the two polarizations of the massless graviton. 

Of particular interest is the $\tilde R^2$ theory in $W_4$ space. Its similar $R^2$ theory in $V_4$ space, has only global scale symmetry with the conserved noether current: $j^N_\mu=2\alpha\phi\nabla_\mu\phi$. On the contrary, the $\tilde R^2$ theory, which is given by the following action: $S=\gamma\int d^4x\sqrt{-g}(\tilde R^2-\beta^2Q^2_{\mu\nu}/2)/2$, where we added the pure nonmetricity term $\propto Q^2_{\mu\nu}$, or, after proper linearization:

\begin{align} S=\frac{1}{2}\int d^4x\sqrt{-g}\left(\frac{\alpha}{6}\phi^2\tilde R-\frac{\lambda}{4}\phi^4-\frac{\beta^2}{2}Q^2_{\mu\nu}\right),\label{w4-r2-action}\end{align} coincides with a particular case of \eqref{w4-action} when $\omega=0$, which is manifestly LS invariant. The derived EOM are:

\begin{align} \alpha\phi^2G_{\mu\nu}&-\alpha(\nabla_\mu\nabla_\nu-g_{\mu\nu}\nabla^2)\phi^2+\frac{3\lambda}{4}\phi^4g_{\mu\nu}\nonumber\\
&-\frac{3\alpha}{2}\phi^2\left(Q_\mu Q_\nu-\frac{1}{2}g_{\mu\nu}Q_\lambda Q^\lambda\right)\nonumber\\
&+3\alpha\left[\der_{(\mu}\phi^2Q_{\nu)}-\frac{1}{2}g_{\mu\nu}\der_\lambda\phi^2Q^\lambda\right]\nonumber\\
&-6\beta^2\left(Q^{\;\;\lambda}_\mu Q_{\nu\lambda}-\frac{1}{4}g_{\mu\nu}Q^2_{\lambda\sigma}\right)=0,\label{w4-r2-eom}\end{align} plus the ``Maxwell'' equation:

\begin{align} \nabla^\nu Q_{\mu\nu}=-\frac{\alpha}{4\beta^2}\left(\phi^2Q_\mu-\der_\mu\phi^2\right),\label{w4-r2-max}\end{align} and the KG equation for the scalar field:

\begin{align} \alpha\phi^2R-\frac{3\alpha}{2}\phi^2Q_\mu Q^\mu-3\alpha\phi^2\nabla_\mu Q^\mu-3\lambda\phi^4=0.\label{w4-r2-kg}\end{align} By comparing this last equation with the trace of \eqref{w4-r2-eom}: $-\alpha\phi^2R+3\alpha\nabla^2\phi^2+\frac{3\alpha}{2}\phi^2Q_\mu Q^\mu-3\alpha\der_\mu\phi^2Q^\mu+3\lambda\phi^4=0,$ we get that $\alpha\nabla_\mu\left(\phi^2Q^\mu-\der^\mu\phi^2\right)=0.$ We can identify the conserved noether current 

\begin{align} j^\mu_N\equiv\alpha\left(\phi^2Q^\mu-\der^\mu\phi^2\right).\label{j-w4}\end{align} This current vanishes only in the particular case when $Q_\mu=\der_\mu\phi^2/\phi^2$, which corresponds to Weyl integrable geometry (WIG). Hence, the quadratic theory \eqref{w4-r2-action} is local scale invariant and possesses a nonvanishing noether current. This result generalizes to $W_4$ space the conclusion made in \cite{oda-2022} on the basis of Riemann geometry. The nonvanishing noether current is due to the gauge transformation of the nonmetricity vector: $Q_\mu\rightarrow Q_\mu-\der_\mu\epsilon(x),$  where the transformation parameter is given by $\epsilon\equiv\ln\Omega^2$.

The study of LS symmetry in gravitational theories is generalized to include the effects of torsion in Sections \ref{sect-ecg}, \ref{sect-ecartan} and \ref{sect-ginv-ech}. In this case the conformal transformation of torsion can not be uniquely established. There is some freedom in the way the torsion transforms under \eqref{weyl-t} as it has been discussed, for instance, in Ref. \cite{shapiro-rev, german-1985, zhytnikov, helayel-neto}. In the case studied in this paper -- SECH theory -- we considered non-propagating (non-dynamical) torsion so that, not new gravitational DOFs arise in addition to those of GR. As a consequence, LS symmetry is accompanied by vanishing of the associated noether current. The result that the torsion is not dynamical is not clear from start. Only after obtaining the solution \eqref{tor-sol}, it is seen that the torsion components are related with the derivatives of the scalar field $\phi$. As long as the SECH theory is LS invariant, since the scalar field is not a dynamical DOF, then the torsion is not dynamical. But if we give up local scale symmetry, then the scalar field becomes a physical DOF and the torsion in the resulting model is dynamical. It would be interesting to extend this study to the case of dynamical torsion \cite{dt-1, dt-2, dt-3, dt-4, dt-5, dt-6, dt-7, dt-8, dt-9} as well. i. e., when there are derivatives of the torsion in the EOM. 

An interesting subject of research can be to look for the effect LS symmetry could play in the so called teleparallel theories \cite{tpt-1, tpt-2, tpt-3, tpt-4, tpt-5, tpt-6, tpt-7, tpt-8, tpt-9, tpt-10, hohmann-1, hohmann-2, hohmann-3, tpt-11, tpt-12, tpt-13}. In Ref. \cite{hohmann-1, hohmann-2, hohmann-3}, for instance, a wide variety of Lagrangians in the context of teleparallel torsion-scalar gravity are investigated. It is shown that the different Lagrangians can be related by conformal transformations of the tetrad and redefinitions of the scalar field. However, the local scale invariance is not studied. In the teleparallel formulation of gravitational theories in $U_4$ space the generalized curvature tensor vanishes: $\tilde R^\alpha_{\;\;\mu\beta\nu}=0$, so that the overall space is flat. The generalized curvature can be decomposed in the following way: $\tilde R^\alpha_{\;\;\mu\beta\nu}=R^\alpha_{\;\;\mu\beta\nu}+\nabla_\beta J^\alpha_{\;\;\nu\mu}-\nabla_\nu J^\alpha_{\;\;\beta\mu}+J^\alpha_{\;\;\beta\lambda}J^\lambda_{\;\;\nu\mu}-J^\alpha_{\;\;\nu\lambda}J^\lambda_{\;\;\beta\mu},$ where $R^\alpha_{\;\;\mu\beta\nu}$, is the Riemann-Christoffel (LC) curvature tensor and $J^\alpha_{\;\;\mu\nu}=L^\alpha_{\;\;\mu\nu}+K^\alpha_{\;\;\mu\nu}$, is the distortion tensor, which amounts to the sum of the disformation \eqref{disf-t} and of the contortion \eqref{contort-t}. The teleparallel geometry may be understood as if riemannian curvature effects were compensated by the effects of nometricity and torsion jointly taken. Since the only way to preserve the teleparallel condition is that the generalized curvature did not transform under the LS transformations, and since under \eqref{weyl-t} the LC connection $\{^\alpha_{\mu\nu}\}$ and the disformation $L^\alpha_{\;\;\mu\nu}$ transform in the same way but with the opposite sign, so that $\{^\alpha_{\mu\nu}\}+L^\alpha_{\;\;\mu\nu}$ does not transform under the Weyl transformations, then LS invariance within the framework of teleparallel gravity theories requires that the torsion (and, consequently, the contortion $K^\alpha_{\;\;\mu\nu}$) did not transform under \eqref{weyl-t}. This type of invariance is known as weak conformal symmetry \cite{shapiro-rev}.

%%%%%%%%%%%%%%%%%%%%%%%%%%%%%%Some prospects%%%%%%%%%%%%%%%%%%%%%%%%%%%%%

Although Riemman-Cartan geometry offers a very wide range of possibilities for conformal symmetry, the most useful transformation of torsion encountered is when in \eqref{conf-t-torsion} we consider the following values of the free parameter: $s=1$ and $s=0$. Even though we considered a generic approach to the possible LS transformations for the torsion, in order to get concrete results we had to restrict our study to the $s=1$ case, as required by LS invariance of the action \eqref{splitedx-action}. One possibility for further work will be to look for a more general class of LS invariant gravitational theories, allowing consideration of generic free constant $s$. Consideration of the boundary terms for the Einstein-Cartan theory is another aspect of Riemann-Cartan geometry that deserves further investigation. The study of these terms may be interesting as they may shed some light on how does the torsion transforms, in order to achieve LS symmetry in proper generalizations of Einstein-Cartan gravity theory. 

Another prospect directly relates to something we mentioned in the introduction. Scale invariance has been recently employed in cosmology in order to model cosmic inflation. The special feature of these models is that they can dynamically break the scale invariance leading to the recovery of General Relativity at the end of inflation. To the best of our knowledge, only global scale invariance has been investigated within the framework of cosmic inflation, therefore, a possible avenue would be to look for the impact of LS symmetry breaking in the primordial inflationary stage of the cosmic expansion.

Despite of the above mentioned possibilities for further work, the natural continuation of the present investigation will be to look for generalization of our results to LS invariant theories under the consideration of the most general non-riemannian geometric framework, consisting of curvature, nonmetricity and torsion at once.

%%%%%%%%%%%%%%%%%%%%%%%%%%%%%%%%%%%%%%%%%%%

\section{Conclusion}\label{sect-conclu}

%%%%%%%%%%%%%%%%%%%%%%%%%%%%%%%%%%%%%%%%%%

In this paper we have generalized previous results of Ref. \cite{jackiw-2015, garay, oda-2022}, that were obtained within the context of Riemann geometry space $V_4$, to non-riemannian geometric framework. Namely, we focused in Weyl geometry space $W_4$ and also in Riemann-Cartan space $U_4$. In \cite{jackiw-2015} it is shown that the conserved noether current associated with LS invariance in CCS theories, vanishes. It is concluded that the corresponding LS symmetry does not have any dynamical role. This result was confirmed within the framework of WTDiff theories in \cite{garay}. Then, in Ref. \cite{oda-2022}, the issue is revisited. Based on the second noether theorem for local symmetry, the authors prove that the noether current associated with the LS symmetry is in general vanishing in any LS invariant gravitational theories in four dimensional riemannian geometry. The reason for this is also clarified: The Weyl transformation is non-dynamical in the sense that it does not contain the derivative term of the transformation parameter: $\der_\mu\epsilon(x)$, as opposed to the conventional gauge transformation.

The most relevant outcome of the present paper is the generalization of the above results to local scale invariant theories of gravity in non-riemannian space. Here we have considered Weyl geometric structure of background space, as well as Riemann-Cartan space with nonvanishing torsion, separately. Consideration of Weyl-Cartan geometric structure is left for further work. In addition we have established a connection between counting of the number of physical DOFs and the dynamical character of local scale symmetry: the noether current associated with LS symmetry vanishes whenever there are not new gravitational DOFs in addition to the two polarizations of the massless graviton. For instance, in LS invariant CCS theory over Weyl geometry space $W_4$, where in addition to the GR gravitational DOFs there are three more propagating DOFs: two transversal and one longitudinal polarizations of the massive nonmetricity vector $Q_\mu$, the noether current is nonvanishing. In the particular case when the nonmetricity vector amounts to the gradient of the compensating scalar field: $Q_\mu=\der_\mu\phi^2/\phi^2$ (properly in WIG space,) since the scalar field is not dynamical, there are only two gravitational degrees of freedom (the two polarizations of the massless graviton.) In consequence, the noether current vanishes, so that LS symmetry is not dynamical in this case. The same is true in the case of the scalar Einstein-Cartan-Holst theory, where the torsion is related with the derivatives of the compensating scalar field. Since the latter is not dynamical, then the torsion is not dynamical either. Therefore, gravity is carried by the two DOFs of the massless graviton as in GR, so that the noether current vanishes in this case as well.

%------------acknowledgments-----------------

{\bf Acknowledgments.} RGQ thanks CONACyT for funding and support. IQ acknowledges FORDECYT-PRONACES-CONACYT for support of the present research under grant CF-MG-2558591.

%%%%%%%%%%%%%%%%%%%%%%%%%%%%%%%%%%%%%%%%%%%%%%
%%%%%%%%%%%%%%%%%%%%%%%%%%%%%%%%%%%%%%%%%%%%%%

\appendix

%%%%%%%%%%%%%%%%%%%%%%%%%%%%%%%%%%%%%%%%%%%%%%

%%%%%%%%%%%%%%%%%%%%%%%%%%%%%%%%%%%%%%%%%%%%%%%%%%%%%%%%%%%%%%%%%%%%%%%%%%%%

\section{Einstein-Cartan theory with the Holst term and variational principle}\label{app-a}

%%%%%%%%%%%%%%%%%%%%%%%%%%%%%%%%%%%%%%%%%%%%%%%%%%%%%%%%%%%%%%%%%%%%%%%%%%%

Let us consider the gauge invariant SECH action \eqref{sech-action} without the $\lambda\phi^4$-term: $S_\text{sech}=\int d^4x L_\text{sech}$, where the Scalar-Einstein-Cartan-Holst Lagrangian is given by:

\begin{align} L_\text{sech}&=\frac{\sqrt{-g}}{2}\left[\phi^2\left(\tilde R+\alpha{\cal R}\right)\right]\nonumber\\
&=\frac{\sqrt{-g}}{2}\left[\phi^2\left(R+{\cal T}+\alpha{\cal R}\right)-2\der_\mu\phi^2T^\mu\right],\label{app-a-1}\end{align} and we have omitted a total derivative. 

Here we shall explore the different variations of this Lagrangian. Let us consider the variation of the scalar-EC term and the scalar-Holst term separately. First, variation of the scalar-EC term with respect to the metric reads:

\begin{align} &\frac{\delta\left(\sqrt{-g}\phi^2\tilde R\right)}{\sqrt{-g}\delta g^{\mu\nu}}=\phi^2G_{\mu\nu}-2\Big[\nabla_{(\mu}\phi^2T_{\nu)}-\frac{1}{2}g_{\mu\nu}\nabla_\lambda\phi^2T^\lambda\Big]\nonumber\\
&-\left(\nabla_\mu\nabla_\nu-g_{\mu\nu}\nabla^2\right)\phi^2+\phi^2\left({\cal T}_{\mu\nu}-\frac{1}{2}g_{\mu\nu}{\cal T}\right),\label{app-a-2}
\end{align} where we have introduced the following quantity:

\begin{align} {\cal T}_{\mu\nu}\equiv&-\frac{1}{4}T_{\mu\sigma\lambda}T_\nu^{\;\;\sigma\lambda}+\frac{1}{2}T_{\lambda\sigma\mu}T^{\lambda\sigma}_{\;\;\;\;\nu}\nonumber\\
&-\frac{1}{2}T_{\lambda\mu\sigma}T^{\sigma\lambda}_{\;\;\;\;\nu}-T_\mu T_\nu,\label{app-a-3}\end{align} whose trace ${\cal T}\equiv g^{\mu\nu}{\cal T}_{\mu\nu}$ is given by \eqref{t-scalar}. 

As an illustration of the procedure followed by us in order to derive the RHS of equation \eqref{app-a-2}, let us consider the variation of the torsion scalar ${\cal T}$, which is defined in \eqref{t-scalar}. First, we split ${\cal T}$ into parts so as to make the derivation more clear:

\begin{align} {\cal T}=\frac{1}{4}{\cal T}_1-\frac{1}{2}{\cal T}_2-{\cal T}_3,\label{app-a-2-1}\end{align} where ${\cal T}_1\equiv T_{\lambda\mu\nu}T^{\lambda\mu\nu}$, ${\cal T}_2\equiv T_{\lambda\mu\nu}T^{\nu\lambda\mu}$ and ${\cal T}_3\equiv T_\mu T^\mu$. Here $T^\mu=g_{\lambda\nu}T^{\lambda\mu\nu}=T^{\lambda\mu}_{\;\;\;\;\lambda}$ while $T_\mu=T^\lambda_{\;\;\mu\lambda}$. We have that:

\begin{align} \delta_{\bf g}{\cal T}_1&=\delta\left(g_{\lambda\sigma}g^{\mu\kappa}g^{\nu\tau}T^\sigma_{\;\;\mu\nu}T^\lambda_{\;\;\kappa\tau}\right)\nonumber\\
 &=\delta g_{\lambda\sigma}T^\sigma_{\;\;\mu\nu}T^{\lambda\mu\nu}+2g^{\mu\kappa}\delta g^{\nu\tau}T_{\lambda\mu\nu}T^\lambda_{\;\;\kappa\tau}\nonumber\\
 &=\left(2T_{\lambda\sigma\mu}T^{\lambda\sigma}_{\;\;\;\;\nu}-T_{\mu\lambda\sigma}T^{\;\;\lambda\sigma}_\nu\right)\delta g^{\mu\nu},\label{app-a-2-2}\end{align} where, in the last line, we took into account that $\delta g_{\lambda\sigma}=-g_{\lambda\mu}g_{\sigma\nu}\delta g^{\mu\nu}$ and, besides, we rearranged indices. Next:

 \begin{align} \delta_{\bf g}{\cal T}_2=\delta\left(g^{\mu\lambda}T^\sigma_{\;\;\mu\nu}T^\nu_{\;\;\sigma\lambda}\right)=T_{\sigma\mu\lambda}T^{\lambda\sigma}_{\;\;\;\;\nu}\delta g^{\mu\nu},\label{app-a-2-3}\end{align} while,

 \begin{align} \delta_{\bf g}{\cal T}_3=\delta\left(g^{\mu\nu}T^\lambda_{\;\;\nu\lambda}T^\sigma_{\;\;\mu\sigma}\right)=T_\mu T_\nu\delta g^{\mu\nu}.\label{app-a-2-4}\end{align} Taking into account \eqref{app-a-2-2}, \eqref{app-a-2-3} and \eqref{app-a-2-4}, we can write:

 \begin{align} \delta_{\bf g}{\cal T}=\frac{1}{4}\delta_{\bf g}{\cal T}_1-\frac{1}{2}\delta_{\bf g}{\cal T}_2-\delta_{\bf g}{\cal T}_3={\cal T}_{\mu\nu}\delta g^{\mu\nu},\label{app-a-2-5}\end{align} where ${\cal T}_{\mu\nu}$ is given by \eqref{app-a-3}.

 Let us now find variation of the Holst term with respect to the metric: $\delta_{g}(\sqrt{-g}\phi^2{\cal R})$. Omitting a total derivative in the Lagrangian \eqref{app-a-1}, we can rewrite the (modified) Nieh-Yan equation \eqref{holst}, in the following way: 

\begin{align}\sqrt{-g}\phi^2{\cal R}=3\sqrt{-g}\der_\lambda\phi^2\hat T^\lambda+\frac{1}{4}\phi^2\epsilon^{\sigma\mu\lambda\nu}T_{\kappa\sigma\mu}T^\kappa_{\;\;\lambda\nu},\nonumber\end{align} or, taking into account the definition of the vector $\hat T^\mu$:

\begin{align} \sqrt{-g}\phi^2{\cal R}=&\frac{1}{2}g_{\mu\nu}\left(\der_\lambda\phi^2\epsilon^{\lambda\mu\sigma\kappa}T^\nu_{\;\;\sigma\kappa}\right.\nonumber\\
&\left.+\frac{1}{2}\phi^2\epsilon^{\sigma\kappa\lambda\theta}T^\mu_{\;\;\sigma\kappa}T^\nu_{\;\;\lambda\theta}\right),\label{app-a-4}\end{align} where, we recall, that $\epsilon^{\sigma\mu\lambda\nu}$ is the totally antisymmetric Levi-Civita symbol ($\epsilon^{0123}=+1$). Then the variation of the Holst term with respect to the metric yields:

\begin{align} \delta_{g}(\sqrt{-g}\phi^2{\cal R})=-\sqrt{-g}\phi^2{\cal R}_{\mu\nu}\delta g^{\mu\nu},\label{app-a-5}\end{align} where we have introduced the ``Holst tensor'' ${\cal R}_{\mu\nu}$ according to the following definition:

\begin{align} {\cal R}_{\mu\nu}\equiv-\frac{1}{2}\frac{\der_\lambda\phi^2}{\phi^2}e_\nu^{\;\;\lambda\sigma\kappa}T_{\mu\sigma\kappa}+\frac{1}{4}e^{\lambda\theta\sigma\kappa}T_{\mu\lambda\theta}T_{\nu\sigma\kappa},\label{app-a-6}\end{align} whose trace is the Holst scalar: ${\cal R}\equiv g^{\mu\nu}{\cal R}_{\mu\nu}$.

In consequence, variation of the SECH Lagrangian \eqref{app-a-1} with respect to the metric yields:

\begin{align} &\frac{2\delta L_\text{sech}}{\sqrt{-g}\delta g^{\mu\nu}}=\phi^2\left(G_{\mu\nu}-\alpha{\cal R}_{\mu\nu}\right)-\left(\nabla_\mu\nabla_\nu-g_{\mu\nu}\nabla^2\right)\phi^2\nonumber\\
&+\phi^2\left({\cal T}_{\mu\nu}-\frac{1}{2}g_{\mu\nu}{\cal T}\right)-2\Big[\nabla_{(\mu}\phi^2T_{\nu)}-\frac{1}{2}g_{\mu\nu}\nabla_\lambda\phi^2T^\lambda\Big],\nonumber\end{align} from where the EOM \eqref{ecart-eom} is obtained.

%%%%%%%%%%%%%%%%%%%%%%%%%%%%%%%

%%%%%%%%%%%%%%%%%%%%%%%%%%%%%

%%%%%%%%%%%%%%%%%

\end{document}